\documentclass[12pt]{article}
\usepackage{amsmath,amssymb,amsfonts}

\begin{document}

\def\Image{\mathop{\text{Im}}}
\def\Ker{\mathop{\text{Ker}}}
\def\tr{\mathop{\text{tr}}}
\def\S{{\text{S}}}

\begin{titlepage}

\title{ \bf Center of quantum group in roots of unity and
the restriction of integrable models\\
{\small\it The talk given in RAQIS03 conference, Aneccy, 25-28/03/03}}

\author {A.~Belavin\\
\small Landau Institute for Theoretical Physics,\\
\small Chernogolovka, 142432, Russia\\
\small e-mail: belavin@itp. ac. ru}

\date{\today}

\end{titlepage}

\maketitle
    
\begin{abstract}
We show the connection between the extended center of the quantum group
in roots of unity and the restriction of the $XXZ$ model.
We also give explicit expressions for operators that respect
the restriction and act on the state space of the restricted models.
The formulas for these operators are verified by explicit calculation
for third-degree roots; they are conjectured to hold in the general case.
\end{abstract}

\section{Introduction}

F.~Alcaraz et al.~\cite{A1} discovered a remarkable fact:
the $XXZ$ model with the special open boundary conditions (OBC) and
a rational value of the anisotropy parameter admits a restriction.
The model arising as a result of the restriction coincides in the
thermodynamic limit with one of the Minimal Models of CFT.
The algebraic reason for the restriction was explained in~\cite{PS}
and~\cite{S}. It was shown in~\cite{PS} that the $XXZ$ model with
the OBC considered in~\cite{A1} has not only integrability but also
$U_q(sl(2))$ symmetry. In roots of unity, the state space of the model
decomposes to the sum of ``good" and ``bad" representations of $U_q(sl(2))$.
The restriction of Alcaraz et al.~is equivalent throwing out ``bad" parts
and keeping only the highest vectors of ``good" representations.
In~\cite{S}, a new monodromy matrix was constructed that is bilinear
in terms of generators of the quantum group $A(u)$, $B(u)$, $C(u)$, $D(u)$
connected in the usual way with the Hamiltonian of the model in~\cite{A1}
and compatible with integrability.
The twisted trace of this monodromy matrix (Sklyanin transfer matrix)
also has $U_q(sl(2))$ symmetry and admits the restriction.

In this paper, we generalize the Pasqier--Saleur construction. We show
that not only $U_q(sl(2))$-invariant but also a much wider class of OBC
indeed admits the restriction (the Sklyanin construction of the transfer
matrix also works for this wider class of OBC). The decisive condition for compatibility
of the Hamiltonian, the transfer matrix, and other operators
with the restiction is their ``weak" commutativity
with a special element of the quantum group.
The notion of ``weak" commutativity and its connection with the extension
of the center of the quantum group in roots of unity is explained below.

In Sec.~2, we describe the conditions that the operators must have
in order to admit the restriction.
In Sec.~3, some such operators are found.
We discuss some possible generalizations
of the construction in the last section.

\section{Center of the quantum group and the restriction in roots of unity}

As usual, let the $R$-matrix $R(u)$ denote the solution
of the Yang--Baxter equation. We consider the simplest and well-known
$R(u)$ matrix corresponding to the six-vertex model,
whose elements can be written as
\begin{equation}
\begin{aligned}
{}&R_{\alpha\alpha}^{\alpha\alpha}(u)=\rho \sin(u+\eta),
\\
{}&R_{\alpha\beta}^{\alpha\beta}(u)=\rho \sin u,
\\
{}&R_{\beta\alpha}^{\alpha\beta}(u)=\rho \sin \eta,
\end{aligned}
\label{8vR}
\end{equation}
where $\alpha,\beta=1,2$, $\alpha\neq\beta$, and
$\eta$ is the so-called anisotropy parameter.

The quantum group {\bf A} connected with $R(u)$ is generated by
$A(u)$, $B(u)$, $C(u)$, $D(u)$, entries of the monodromy matrix $L(u)$,
which satisfies
\begin{equation}
R_{12}(u-v)L_1(u)L_2(v)=L_2(v)L_1(u)R_{12}(u-v).
\label{YBA}
\end{equation}
As shown by V.~Tarasov~\cite{T}, the center of {\bf A} in roots of unity,
i.e., $\eta=\pi m/N$, is generated by the following elements of {\bf A}:
$$
\begin{aligned}
{}&\langle A(u)\rangle =A(u)A(u+\eta)\cdots A(u+(N-1)\eta),
\\
{}&\langle B(u)\rangle =B(u)B(u+\eta)\cdots B(u+(N-1)\eta),
\\
{}&\langle C(u)\rangle =C(u)C(u+\eta)\cdots C(u+(N-1)\eta),
\\
{}&\langle D(u)\rangle =D(u)D(u+\eta)\cdots D(u+(N-1)\eta).
\end{aligned}
$$

For convenience, we let $\langle B(u)\rangle $ denote the central element,
$$
\langle B(u)\rangle =B(u)B_1(u),
$$
where
$$
B_1(u)= B(u+\eta)\cdots B(u+(N-1)\eta).
$$
We now fix $V= C^2\otimes\cdots\otimes C^2$ as the representation
space of our quantum group.
It is easy to see that for arbitrary $v$,
$$
\langle B(v)\rangle =B(v)B_1(v)=0
$$
on this space.
We can then define the state space of a restricted model as
$$
W(v)=\Ker B(v)/\Image B_1(v).
$$

In the limit $v\to\infty$, $B(v)$ coincides up to a scalar
factor with $X$, one of the generators of $U_q(sl(2))$. As a result,
$W(\infty)$ coincides with the space of the ``good" highest vectors
of Pasqier--Saleur.
It was shown in~\cite{PS} that the Hamiltonian of the $XXZ$ chain
with OBC of special type
$$
H_{XXZ} = \sum^{L-1}_{n=1}\biggl[\sigma^+_n\sigma^-_{n+1}
+\sigma^-_n\sigma^+_{n+1}
+\frac{\cos\eta}{2}\sigma^z_n\sigma^z_{n+1}
+i\frac{\sin\eta}{2}(\sigma^z_n - \sigma^z_{n+1})\biggr]
$$
is invariant under the quantum algebra $U_q(sl(2))$. Here, $q=e^{\eta}$.
Because of this, $H_{XXZ}$ is properly defined
on $W(\infty)=\Ker X/\Image X^{(N-1)}$.

In the thermodynamic limit, where $L\to\infty$, the spectrum of low-lying
states coincides (in Cardy's sense) with $M(N-1/N)$, one
of the Minimal Models of CFT.
In the next section, we show that the construction in~\cite{PS} can be
generalized to arbitrary values of the parameter $v$.
The corresponding Hamiltonian is~\cite{S},~\cite{A2}
\begin{align}
H_{XXZ}={}&\sum^{L-1}_{n=1}\biggl[\sigma^+_n\sigma^-_{n+1}
+\sigma^-_n\sigma^+_{n+1}
+\frac{\cos\eta}{2}\sigma^z_n\sigma^z_{n+1}\biggr]
\nonumber
\\
&{}+i\frac{\sin\eta}{2}(\cot v \sigma^z_1 -\cot(v+\eta) \sigma^z_L).
\label{H}
\end{align}

We now find the sufficient conditions for any operator $Q$ to be projectible
on $W(v)$. It is easy to see that sufficient conditions are that there exist
some operators $\widehat Q$ and $\widehat Q_1$ for a given $Q$ such that
\begin{equation}
\label{A}
B(v)Q=\widehat Q B(v)
\end{equation}
and
\begin{equation}
\label{B}
Q B_1(v)=B_1(v) \widehat Q_1.
\end{equation}
Indeed, Eq.~\eqref{A} guarantees that if a vector $\psi\in\Ker B(v)$,
then the vector $Q\psi\in\Ker B(v)$.
It follows from Eq.~\eqref{B} that if the difference of two vectors
$\psi_1$ and $\psi_2$ belong to $\Image B_1(v)$,
i.e., if $\psi_1-\psi_2=B_1(v)\chi$,
then the difference of $Q\psi_1$ and $Q\psi_2$ also belongs to $\Image B_1(v)$.

\section{Sklyanin transfer matrix and other operators
respecting the restriction in roots of unity}

In~\cite{S}, E.~Sklyanin explained the integrability of the $XXZ$ model
with OBC of the form
\begin{align*}
H_{XXZ}={}&\sum^{L-1}_{n=1}\biggl[\sigma^+_n\sigma^-_{n+1}
+\sigma^-_n\sigma^+_{n+1}
+\frac{\cos\eta}{2}\sigma^z_n\sigma^z_{n+1}\biggr]
\\
&{}+i\frac{\sin\eta}{2}(\cot(\xi_++\eta/2)\sigma^z_1+\cot(\xi_--\eta/2)\sigma^z_L).
\end{align*}
by constructing a special monodromy matrix and using it
to diagonalize Hamiltonian~\eqref{H} and the corresponding transfer matrix by
means of the algebraic Bethe ansatz.

Let $ K_+(u)=K(u+\eta/2,\xi_+)$ and $K_-(u)=K(u-\eta/2,\xi_-)$, where
$$
K(u,\xi)=
\left[\begin{matrix}\sin(u+\xi)&0\\0&-\sin(u-\xi)\end{matrix}\right].
$$
Then $K_{\pm}$ satisfies the boundary Yang--Baxter
equations~\cite{S},~\cite{W}.

The Sklyanin monodromy matrix is defined~\cite{S} as
\begin{align}
\Lambda(u)&=\sigma^2L^t(-u)\sigma^2K_+(u)L(u)
\nonumber
\\
&=\left[\begin{matrix}\Lambda_1^1&\Lambda_1^2\\
\Lambda_2^1&\Lambda_2^2\end{matrix}\right].
\nonumber
\end{align}
Using~\eqref{YBA} and the boundary Yang--Baxter equation,
we can prove that $\Lambda(u)$ satisfies the same equation as $K_+(u)$.
It gives the commutation relations between $\Lambda^i_j$.
The Sklyanin transfer matrix is defined as
\begin{align}
T_\S(u)&=\tr\Lambda(u)K_-(u)
\nonumber
\\
&=\sin(u-\eta/2+\xi_-)\Lambda^1_1-\sin(u-\eta/2-\xi_-)\Lambda^2_2.
\nonumber
\end{align}

Explicit expressions for the elements of the Sklyanin monodromy matrix are
\begin{align}
&\Lambda^1_1(u)=\sin(u+\eta/2+\xi_+)A(u)D(-u)+
\sin(u+\eta/2-\xi_+)C(u)B(-u),
\nonumber
\\
&\Lambda^2_2(u)=-\sin(u+\eta/2+\xi_+)B(u)C(-u)-
\sin(u+\eta/2-\xi_+)D(u)A(-u),
\nonumber
\\
&\Lambda^2_1(u)=\sin(u+\eta/2+\xi_+)B(u)D(-u)+
\sin(u+\eta/2-\xi_+)D(u)B(-u),
\nonumber
\\
&\Lambda^1_2(u)=-\sin(u+\eta/2+\xi_+)A(u)C(-u)-
\sin(u+\eta/2-\xi_+)C(u)A(-u).
\nonumber
\end{align}

Let $\xi_+=v-\eta/2$ and $\xi_-=-v-\eta/2$. Then the following
relations are satisfied:
\begin{equation}
\label{1}
B(v)T_\S(u)=\widehat T_\S(u)B(v),
\end{equation}
where the explicit expression for the Sklyanin transfer matrix for these
values $\xi_{pm}$ is
\begin{align}
T_\S(u)={}&\sin(u-v-\eta)\sin(u+v)A(u)D(-u)
\nonumber
\\
&{}+\sin(u-v-\eta)\sin(u-v+\eta)C(u)B(-u)
\nonumber
\\
&{}+\sin(u+v)\sin(u+v)B(u)C(-u)
\nonumber
\\
&{}+\sin(u+v)\sin(u-v+\eta)D(u)A(-u)
\label{SK}
\end{align}
and
\begin{align*}
\widehat T_\S(u)={}&\sin(u-v)\sin(u+v+\eta)A(u)D(-u)
\\
&{}+\sin(u-v)\sin(u-v)C(u)B(-u)
\\
&{}+\sin(u+v+\eta)\sin(u+v-\eta)B(u)C(-u)
\\
&{}+\sin(u+v+\eta)\sin(u-v)D(u)A(-u).
\end{align*}
We have
\begin{equation}
\label{2}
B(v)\Lambda^2_1(u)=\hat \Lambda^2_1(u)B(v),
\end{equation}
where explicitly
\begin{align*}
&\Lambda^2_1(u)=\sin(u+v)B(u)D(-u)+\sin(u-v+\eta)D(u)B(-u),
\\
&\hat\Lambda^2_1(u)=\sin(u+v+\eta)B(u)D(-u)+\sin(u-v)D(u)B(-u).
\end{align*}
Equations~\eqref{1} and~\eqref{2} were verified by direct calculation.

The operators $T_\S(u)$ and $\Lambda^2_1(u)$ thus satisfy the first
condition, Eq.~\eqref{A}, for the restriction. We conjecture
that they also satisfy the second one, Eq.~\eqref{B}, if $q$ is a root
of unity. This conjecture was explicitly verified by direct calculation
for third-degree roots for the case $T_\S(u)$ (but not for the
case $\Lambda^2_1(u)$). It would nice to find an elegant general proof.

The operators $T_\S(u)$ and $\Lambda^2_1(u)$ depend on one parameter.
There also exists a two-parameter family of operators satisfying~\eqref{A}
and~\eqref{B}.

By definition, let
\begin{align*}
T(x_{ij};u_1,u_2)={}&x_{11}A(u_1)D(u_2)+x_{22}D(u_1)A(u_2)
\\
&{}+x_{12}B(u_1)C(u_2)+x_{21}C(u_1)B(u_2).
\end{align*}
Then
\begin{equation}
\label{3}
B(v)T(x_{ij}(v);u_1,u_2)=T(\hat x_{ij}(v);u_1,u_2)B(v),
\end{equation}
where $T(x_{ij}(v);u_1,u_2)$ and $T(\hat x_{ij}(v);u_1,u_2)$
are obtained from $T(x_{ij};u_1,u_2)$ by suitably substituting
$x_{ij}(v)$ and $\hat x_{ij}(v)$ for $x_{ij}$ and
\begin{align*}
&x_{11}(v)=\sin(u_1-\eta-v)\sin(u_2-v),
\\
&x_{22}(v)=\sin(u_1+\eta-v)\sin(u_2-v),
\\
&x_{12}(v)=-\sin(u_2-v)\sin(u_2-v),
\\
&x_{21}(v)=-\sin(u_1+\eta-v)\sin(u_1-\eta-v),
\\
&\hat x_{11}(v)=\sin(u_1-v)\sin(u_2-\eta-v),
\\
&\hat x_{22}(v)=\sin(u_1-v)\sin(u_2+\eta-v),
\\
&\hat x_{12}(v)=-\sin(u_2-\eta-v)\sin(u_2+\eta-v),
\\
&\hat x_{21}(v)=-\sin(u_1-v)\sin(u_1-v).
\end{align*}
We have
\begin{align}
&T( x'_{ij}(v);u_1,u_2)B(v+\eta)B(v+2\eta)
\nonumber
\\
&\qquad\qquad\qquad\qquad
=B(v+\eta)B(v+2\eta)T( \hat x'_{ij}(v);u_1,u_2),
\label{4}
\end{align}
where
\begin{align*}
&x'_{11}(v)=\sin(u_1-\eta-v)\sin(u_2+3\eta-v),
\\
&x'_{22}(v)=\sin(u_1-2\eta-v)\sin(u_2-v),
\\
&x'_{12}(v)=-\sin(u_2-v)\sin(u_2-3\eta-v),
\\
&x'_{21}(v)=-\sin(u_1-\eta-v)\sin(u_1-2\eta-v),
\\
&\hat x'_{11}(v)=\sin(u_1-3\eta-v)\sin(u_2-\eta-v),
\\
&\hat x'_{22}(v)=\sin(u_1-v)\sin(u_2-2\eta-v),
\\
&\hat x'_{12}(v)=-\sin(u_2-\eta-v)\sin(u_2-2\eta-v),
\\
&\hat x'_{21}(v)=-\sin(u_1-v)\sin(u_1-3\eta-v).
\end{align*}

If we require that the operators $T( x_{ij}(v);u_1,u_2)$ and
$T( x'_{ij}(v);u_1,u_2)$ coincide, we can verify that
this requirement is satisfied if $\eta=\pi/3$ or $\eta=2\pi/3$.
As discussed above, it follows that the two-parameter
family of operators $T( x_{ij}(v);u_1,u_2)$ can be restricted
on $W(v)$.

We conjecture that $T(x_{ij}(v);u_1,u_2)$ satisfies restriction
conditions~\eqref{A} and~\eqref{B} if $\eta=m\pi/p$,
where $m$ and $p$ are coprime integers.

It is easy to verify that the relation
$$
T_\S(u)=T(x_{ij}(v);u,-u)
$$
holds, where $T_\S(u)$ is Sklyanin transfer matrix in~\eqref{SK}.

\section{Discussion}

It was shown in~\cite{BS} that the Sklyanin transfer matrix
for the Pasqier--Saleur case ($v\to\infty$) after the restriction
satisfies the truncated system of fusion functional equations.
This system defines the spectrum $M(p/p+1)$.
This statement can also be generalized to finite $v$.
It is remarkable that the spectrum of states
surviving after the restriction
is independent of $v$~\cite{B}.
This fact was discovered numerically in~\cite{A2}.

The explicit construction for additional central elements
of the elliptic Yang--Baxter algebra in roots of unity
was given in~\cite{BJ}.
It would be interesting to generalize the approach in this
paper to the elliptic case.

Another important problem is to generalize
the Kitanine--Maillet--Terras construction~\cite{Kit}
of the local operators of the $XXZ$ model in
terms of elements of monodromy matrix to the restricted models.
Namely, it would be interesting
to build explicit operators that simultaneously respect
the restriction and have mutual locality (i.e., commutativity).
Constructing such operators would allow obtaining
explicit formulas for the correlation functions
in the restricted models.

\subsection*{Acknowledgments}
I am indebted to M.~Jimbo, N.~Kitanine, and E.~Sklyanin
for the useful discussions and also to W.~Everett
for the editorial assistance.
This work was presented at the conference
``On Recent Advances in the Theory of Quantum Integrable Systems."
I indebted to the organizers of the conference and especially
to P.~Sorba for the opportunity to participate in this nice and very
interesting meeting. This work is supported in part by RFBR-01-02-16686,
SSRF-20044.2003.2, and INTAS-00-00055.

\end{document}